\documentclass[mathleft,fleqn,%
]{an}
\usepackage{graphicx}
\usepackage[varg]{txfonts}
\begin{document}

\Pagespan{1}{}
\Yearpublication{2014}%
\Yearsubmission{2014}%
\Month{0}%
\Volume{999}%
\Issue{0}%
\DOI{asna.201400000}%

\title{Chaos Control with Ion Propulsion}

\author{J. Sl\'iz \inst{1}\fnmsep\thanks{Corresponding author:
        {sliz.judit@astro.elte.hu}},
 T. Kov\'acs \inst{2,3}
\and \'A. S\"uli \inst{1,3}
}
\titlerunning{Instructions for authors}
\authorrunning{T.\,H.\,E. Editor \& G.\,H. Ostwriter}
\institute{
Department of Astronomy, E\"otv\"os University, P\'azm\'any P. s. 1A , Budapest, 1117, Hungary
\and 
Institute for Theoretical Physics, E\"otv\"os University, P\'azm\'any P. s. 1A , Budapest, 1117, Hungary
\and 
Institute for Astronomy and Earth Sciences, Hungarian Academy of Sciences, P.O. 67, H-1125, Budapest, Hungary}

\received{XXXX}
\accepted{XXXX}
\publonline{XXXX}


\abstract{The escape dynamics around the triangular Lagrangian point
 $L_5$ in the real Sun--Earth--Moon--Spacecraft system is
  investigated. Appearance of the finite time chaotic behaviour
  suggests that widely used methods and concepts of dynamical system
  theory can be useful   in constructing a desired mission design. Existing
  chaos control methods are modified in such a way that we are able to
  protect a test particle from escape. We introduce initial condition
  maps in order to have a suitable numerical method to describe the
  motion in high dimensional phase space. Results show that the
  structure of initial condition maps can be split into two
  well-defined domains. One of these two parts has a regular
  contiguous shape and is responsible for long time escape;  it is a
  long-lived island. The other one shows a filamentary fractal
  structure in initial condition maps. The short time escape is
  governed by this object. This study focuses  on a low-cost method
  which successfully transfers a reference trajectory between these
  two regions using an appropriate continuous control force. A
  comparison of the Earth--Moon transfer is also presented to show the
  efficiency of our method.}

\maketitle

\section{Introduction}
Finite time chaotic behaviour is a common phenomenon in simple as well as
in complex dynamical systems. Transient chaos in Hamiltonian
dynamics appears as chaotic scattering (Bleher, Grebogi \& Ott 1990)   or half-scattering. This
can later be thought as of escape or capture from or into a
system. The finite time chaotic motion might proceed before the
particle leaves the systems, say interacting with a scatter object,
after which the irregular motion might cease. 

Escape in simple dynamical systems has been extensively studied  for
roughly 25 years, {\it for more details Ott (1993), Gaspard (1998), Contopulos (2004), Lai \& T\'el (2011)}. Theory of transient chaos states that finite time
chaotic behaviour is related to the unstable periodic orbits and their
stable and unstable manifolds in the phase space. In addition, one can construct
numerically a well-defined fractal set in the phase space containing
all the heteroclinic and homoclinic intersections of these manifolds 
 as well as the unstable periodic orbits themselves. This fractal set is 
responsible for finite time chaotic behaviour. Some textbooks call
this set the chaotic saddle, indicating its non-attracting nature. 

The large number of studies dealing with escape dynamics shows the importance of
this issue in dynamical astronomy. From the very simple systems such as
the Hill approximation of 
the restricted three body problem(Sim\'o \& Stuchi 2000) or the Sitnikov problem (Kov\'acs \& \'Erdi 2009) to more
complex dynamics, e.g. 3D model of barred galaxies (Contopoulos \& Harsoula 2013; Jung \& Zotos 2015; Maffione et al. 2015) or open star
clusters (Ernst et al. 2015), one can find thorough investigations of finite time chaotic motion. 
Numerical simulations and also analytic computations  have  pointed out
that the stable and unstable manifolds of 
the unstable periodic orbits play a crucial role in escape and 
capture (Ashtakov et al. 2004; Branicki \& Wiggins 2010; Harsoula, Contopoulos \& Efthymiopoulos 2015;  Efthymiopoulos 2012).
In computing the unstable periodic orbits and their manifolds, two
suitable analytical methods are commonly used, one of these two is called the
hyperbolic normal form (Moser 1958),  and the other method is the normally
hyperbolic invariant manifolds, {\it introduced by Fenichel (1979)}.  
The chaotic saddle can also be constructed numerically by using the
sprinkler method (T\'el \& Gruiz 2006).

At the end of the last century,  researchers and engineers started to use
the manifolds of periodic orbits to describe the dynamics of
spacecraft  manoeuvres and station keeping.  The first method of constructing trajectories from the Earth to the Moon with lunar ballistic capture using low thrust spiral orbit  {\it was  published by Belbruno (1987).}  Determining the weak stability boundary  around  the Moon was a great leap forward  in the study of the low thrust orbits (Kevin, Belbruno \& Topputo 2012). Although they used  methods similar to ours, but without the conception of transient chaos (survival probability, escape rate, chaotic saddle, etc.) they did not give the explication to the phenomena they experienced (e.g. the magnitude of the thrusting, the stabilization of the motion). 

The new dynamical system
approach technique seemed to be  an  extremely useful application in space research,
especially for precise orbit determination. Many  recent space
missions (Genesis (Howell et al. 1998), MAP (Bennett 1996)  and Herschel-Planck (Rosa et al. 2005)) were
designed  using this so-called space manifold dynamics method (SMD) (Bell\'o, G\'omez \& Masdemont 2010; Belbruno 2010).

The basic purpose of this technique is to find
an orbit from one place to other, or keep a spacecraft at a
given position around an unstable periodic orbit as cheaply as possible,
i.e. with low energy/fuel requirements. SMD  usually involves  the
approximation of the circular restricted problem of three bodies, say, for
instance Sun-Earth-Spacecraft or Earth-Moon-Spacecraft systems,
depending on current application.  

There are five equilibrium points  in an orbital configuration of two large bodies (in our case the Earth and the Moon), where the forces of gravity balance out;  these are the  Lagrangian points. There are five such points,  labelled $L_1$ to $L_5$. The $L_1$, $L_2$ and $L_3$ (collinear) points are unstable, the $L_4$ and $L_5$ (triangular) points are linearly stable equilibrium points. The presence of a third large body, the Sun,  destroys the stability of the Earth-Moon  $L_4$ and $L_5$ points, but  even though these points ($L_4$ and $L_5$) are unstable, there are bounded regions around these points for either shorter or longer periods of time. 

Although the attention in space manifold dynamics has been focused
mainly in collinear points $L_1$, $L_2$, and $L_3$   because of their unstable nature (Celletti \& Giorgili 1991)
and vicinity to the Earth ($L_1$, $L_2$). It can also be shown that escape
from the vicinity of a natively stable libration point (e.g. $L_4$ or $L_5$
in the circular restricted three body problem (CRTBP)) is also governed by the unstable manifold of the chaotic
saddle (Sl\'iz, S\"uli \& Kov\'acs 2015) which encloses the regular islands surrounding the stable
periodic orbit. This domain of the phase space, where the saddle and
the remnants of the outermost KAM torus form a "fat" fractal set, is
responsible for the stickiness effect, one of the fundamental
phenomena in Hamiltonian escape dynamics with mixed phase
space (Contopoulos \& Harsoula 2008; T\'el \& Lai 2008).

We started  test particles  close to  the triangular libration
point $L_5$ of the restricted three body problem (RTBP) and analyzed  their short and long time dynamics  in the presence of a third large body (the Sun), (RFBP). 

As it has already been shown (\'Erdi et al. 2007; Freistetter 2006; Schwarz et al. 2009), the size and the position of the
regular domain around the triangular points depend on several
parameters such as the mass, eccentricity and the orbital position.

Starting a test particle from
the vicinity of the $L_4$, $L_5$ points in the RFBP is not a guarantee that
this trajectory will be stable for  a long time  or that  the particle  will remain
in the system. This is a consequence  deriving from the
general fact in Hamiltonian dynamics that the size of the regular tori in
phase space decays exponentially with the degree of freedom of the
system (Lichtenberg \& Liebermann 1983; Altmann \& Kantz 2007). 

Our primary goal is to avoid the escape and to keep the test particle
in a finite volume of the configuration space with minimal energy
consumption. To do this we use the basic idea of chaos control 
where the main argument is that small changes along a chaotic orbit
should be efficient enough to reach a desired performance of the system
according to some  criteria. The focus is on the  expression  "\textit{small changes}" that can be
applied with success, since the chaos is naturally sensitive to small
changes and therefore tiny perturbations are suitable for control and
manipulation of the chaotic process (Boccaletti \& Pucacco 2000). In this
study, a slightly modified chaos control method is used  which is suitable for our purposes.
We construct the stable manifold of the chaotic 
saddle numerically for a given state of the phase
space, called initial condition 
map (ICM) (Utku, Hagen \& Palmer 2015) and then try to adjust the test particle's velocity
(and consequently the position) to be as close as possible to the
stable manifold.  

The paper is organized as follows: In Section 2 we give
a description of the model and details of methods we used. Our results
are presented in the next section, Section 3. Finally,
in Section 4 we summarize our findings and draw 
conclusions. 

\section{Model and Methods}

\begin{figure}
	
	\includegraphics[width=\columnwidth]{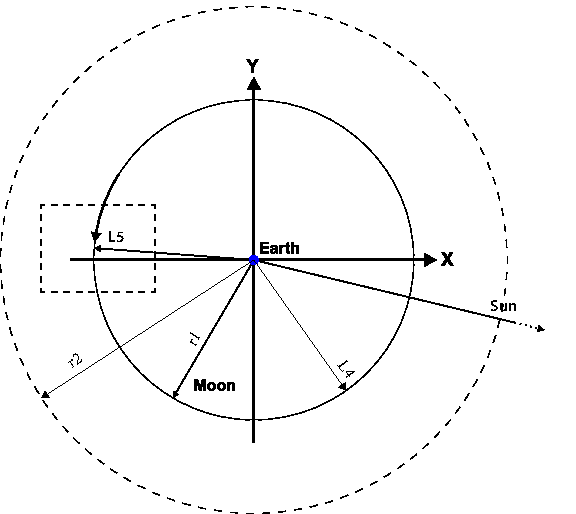}
    \caption{Schema of the system configuration  centred at the Earth. The position
      of the  triangular point $L_5$, and the Moon  are at the correct scale, however, the Sun's position is indicated by an arrow. The rectangle depicts the domain $\mathcal{D}$ where initial conditions are placed. All the quantities are  dimensionless, the characteristic length unit is  $\mathrm{4\times10^5}$ km.  The dimensionless nominal distance of the Moon from the Earth  is  $r_1=1$. The dashed circle $r_2=\sqrt{x^2+y^2}=1.5$ around the Earth indicates the region within which we want to keep the spacecraft.}
    \label{fig:coosys}
\end{figure}

Investigating the escape dynamics, we use a real planar RFBP consisting  of three massive bodies, the primaries,
and a mass-less test particle. To have a realistic application, the actual
parameters of the massive objects in our own solar system are
chosen. That is, the positions and velocities of the Sun, Earth, and
Moon are taken from the freely available NASA JPL database\footnote{http://ssd.jpl.nasa.gov/horizons.cgi} for
the epoch JD = 2455999.5 (2012 March 13 00:00 UT), see
Table~\ref{table:epoch}. Since we investigate the Earth - Moon Lagrangian points, for the sake of better understanding we used a 2D geocentric model. A  coordinate transformation is performed 
to have the originally in inertial frame described equations of motion in the geocentric system
(Fig.~\ref{fig:coosys}). The equations are implemented in
dimensionless form where the characteristic length unit is
$\mathrm{4\times10^5}$ km, and the time unit is 1 day. The computation
method is an adaptive step size Runge-Kutta-Fehlberg 7(8)  integrator (Fehlberg 1968)
in which the actual step size is determined according to the desired
accuracy, $\epsilon=10^{-16}.$ We also set the mass of the test
particle at 1 kg in order to be able to compute the energy consumption with a real spacecraft (Section 3.4). The reason for the choice the RFBP
is to illustrate the Sun's effect to the size, position and
essentially the escape dynamics from the regular domain around the $L_5$
Lagrangian point. Since the Sun is included as a perturber to the
Earth--Moon--test particle three body problem, the escape from the
vicinity of a triangular Lagrangian point is more probable than in the
case without the Sun. 

\begin{table}
\caption{ The initial geocentric conditions of the Sun, Moon, and test particle taken from the JPL database. For units, see text.}
\begin{tabular}{lllll}\hline
& Sun & Moon & test particle\\
&&&close to  $L_5$\\
 &&&(RTBP)\\\hline
$x_0$ &368.666440265 & -0.51661666298 & -0.91418201074\\
$y_0$ &-47.600836868 & -0.75733769053 &  0.06873430889\\
$v_{x0}$ &0.93068574512 &  0.1853827247 &  -0.02527332186\\
$v_{y0}$ &6.40398916643 & -0.13621388441 & -0.22865309127\\\hline
\end{tabular}
\label{table:epoch}
\end{table}

In order to monitor the finite time chaotic behaviour and escape
dynamics,  we place a large number ($\mathrm{9\times10^4}$ to $\mathrm{1\times10^6}$) of particles around the $L_5$ point in
the RFBP, see rectangle in Fig.~\ref{fig:coosys}. All particles start
with the same velocity ($v_{x0},\;v_{y0}$) which is equal to the
velocity of the $L_5$ point in the circular RTBP. 

As is mentioned in Section 1, we want to  keep a test
particle in a pre-defined region of configuration space using a
 suitably designed chaos control method. Most of  the available chaos
control methods,  either feedback or delayed, use the fact that there
is an accessible system parameter that can be adjusted. However, in our
particular problem there are no such parameters. One can only vary the
system states, for example the velocity of the spacecraft, during the
motion. Therefore, we briefly summarize the basic idea of  the two chaos control
methods (Ott, Grebogi \& York 1990; Lai 1996) we used, and then review what changes we
applied to have a suitable control procedure. 

The basic idea of OGY (from the name of Ott, Grebogi and York) method is to identify the unstable low-periodic
orbits in the phase space embedded into a chaotic set (an attractor in
dissipative case) and then applying a  small control on a system
parameter in such a way that the desired periodic orbit would be
stabilized. The stabilization in this sense means practically to move
the trajectory as close as possible to the stable manifold of the
chosen unstable periodic orbit (UPO). The other method, {\it proposed by Lai (1996)},  deals with
targeting and controlling fractal basin boundaries. This method uses
small feedback control to drive trajectories to a pre-selected
attractor.  In performing the small perturbation to the system, one needs
to have initial conditions close to the basin boundaries. This means that
the magnitude of the control depends on how far the initial conditions
lie from the desirable basin boundary. In other words, if there is a
large contiguous domain associated to an undesirable attractor and the
trajectory originates well inside this region, no small controls can
be applied to transfer the trajectory to the other (desirable)
basin. Consequently, small perturbations lead to success,  which are more
likely in cases when the basin boundaries are fractal sets (box
counting dimension $D<1$), which is particularly common in high
dimensional chaotic systems. 

In what follows, we describe how to use the above mentioned methods to
the RFBP.  Although there are many parameters (masses, eccentricities of primaries) 
in a real situation we are not able to vary any of them. Therefore, we might want to adjust the
velocity of the test particle in order to control its phase space
position. This can be done by using a tangential force continuous in time. 
Initial condition maps can be 
suitable representations of the dynamics that tell us how to apply the control force.
A brief description of the intervention procedure is given as follows:
\begin{enumerate}
\item \label{itm:r} a pre-defined region  which we are interested in as a
  possible domain of controlled motion, $r<1.5$  is chosen  around the Earth (Figure~\ref{fig:coosys}). If
  the trajectory leaves this part of the configuration space, it is
  recorded as an escaped trajectory;
\item \label{itm:N0} the motion of large number of particles originating around the $L_5$ point $(x_{0},y_{0})\in \mathcal{D}$ (including that one we want to control, i.e. the reference trajectory (RT)) is integrated;
\item \label{itm:T} a specific integration time, $T,$ is set as long as the
  ensemble is followed. Hence, we can store the $x_0$ and $y_0$
  initial positions of the trajectories, associated with given $v_{x0}$ and $v_{y0},$ surviving until $T.$ See  example Fig.~\ref{fig:rtbpl5} -- this procedure serves a well-defined ICM;
\item \label{itm:SM} we follow the dynamics of the particles until the
  velocity of the RT fulfils the condition $v_x=0,\;v_y<0$ then
  repeat step~\ref{itm:T}, i.e. we start the ensemble again from
  $\mathcal{D}$ but with the updated initial velocities ($0,v_{y}$).
\end{enumerate}

Step~\ref{itm:SM} provides a series of the ICMs for those trajectories
which start from domain $\mathcal{D}$ (Fig.~\ref{fig:coosys}) with velocity $(v_x=0,\;v_y<0).$ 
The points drawing these maps correspond to the stable manifold of the chaotic saddle
in the $x-y$ section of the phase space. In other words, trajectories
on these ICMs do not escape the system sooner than the integration time $T.$

From another point of view, in Hamiltonian systems with escape, one
can consider that the system has an attractor at infinity.  This means,
there is a set of points on the ICMs whose end state is 
infinity, or specifically in our case $r>$1.5 (white region in Figure~\ref{fig:rtbpl5}). Consequently, the
disjointed set of this basin of attraction (black structure in Fig.~\ref{fig:rtbpl5}) yields the trajectories in the ICM not escaping before time $T.$  Therefore, if we wait to keep the test particle at least until $T$ in the system, we should move it away from the infinity's basin boundary. 

\section{Results}
{\it Lissauer \& Chambers (2008); Sl\'iz , S\"uli \& Kov\'acs (2015); P\'aez \& Efthymiopoulos (2015)} showed that particles relative close to the $L_5$  Lagrangian point may leave the system both in RTBP and RFBP. The dynamics of the escaping trajectories can be described by the well-known formalism of transient chaos (Lai \& T\'el 2011).

\subsection{ The Effect of the Sun's Mass}
\begin{figure}
	\includegraphics[width=\columnwidth]{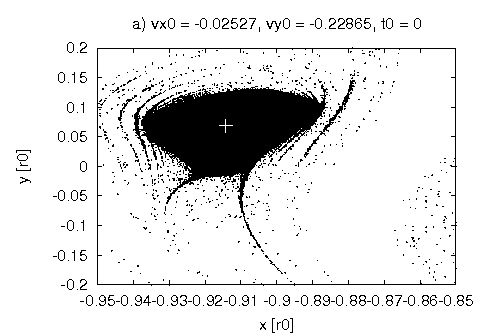}
	\includegraphics[width=\columnwidth]{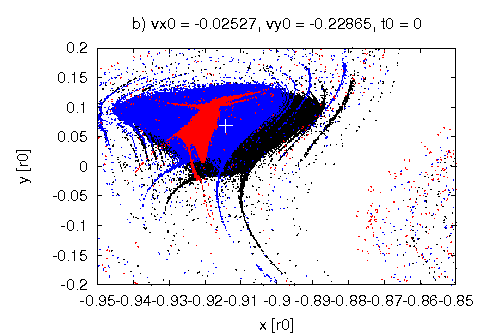}
    \caption{ICMs for 1300 days. (a) Black dots represent the initial
      conditions ($x_{0},y_{0}$) for 90000 trajectories that do not
      escape the system (RTBP) until $T.$ (b) Non-escaped trajectories
      for  the different mass of the Sun. Black, blue, and red dots correspond
      to 0 (i.e. RTBP), 0.6, and 1 solar mass, respectively. Initial
      velocities in both panels correspond to the velocity of the $L_5$
      point in RTBP.  The white cross denotes the position of the triangular Lagrangian point $L_5$.}
    \label{fig:rtbpl5}
\end{figure}

First we demonstrate the effect of the Sun's gravitational
perturbation on the size and position of the domain containing
non-escaping trajectories for $T=1300$ days ( $\sim 6.5\tau$), see
Fig.~\ref{fig:rtbpl5}.  Panel (a) depicts the (ICM) ($x_0,y_0$) for RTBP in region $x_0\in[-0.95;-0.85],$
$y_0\in[-0.2;0.2].$ Initial velocities for all trajectories equal to the
Keplerian velocity of the $L_5$ point, see Table~\ref{table:epoch}. As
is expected from classical celestial mechanics, the triangular
Lagrangian point (white cross) is situated deep inside the area of
long-lived trajectories. The filamentary structure around the extended
region indicates the chaotic motion of escaping particles.

Fig.~\ref{fig:rtbpl5} (b) shows the ICM for different solar 
masses ($M_{\odot}$). Black dots represent $M_{\mathrm{Sun}}=0,$  e.g.  RTBP,
blue dots $M_{\mathrm{Sun}}=0.6 M_{\odot},$ and red dots $M_{\mathrm{Sun}}=1 M_{\odot}.$
Two well-seen consequences can be
observed due to the presence of the Sun. Firstly, the size of the domain of non-escaping trajectories
diminishes  in  case of the
larger Sun's mass. Secondly, the position of this area is shifted
outward from Earth. As a consequence of this drift, the location of the $L_5$ point relative to the stable island moves toward its border and beyond. We note that the
position of the equilibrium point in the RFBP is situated at a different position as in the RTBP
(Gabern \& Jorba 2001). Considering this fact, it is worth monitoring the fate of the
particle starting from the $L_5$ point ($x_0=-0.914,\;y_0=0.068,
v_{x0}=-0.025,\;v_{y0}=-0.228$) and its neighbourhood.


\subsection{The Effect of the Integration Time}
In this section the size of the long-lived island is investigated for
different integration times. That is, we are interested in the number
(and also the position) of non-escaped trajectories, while the
criterion for escape ($r>1.5$) remains the same. The mass of the Sun
is set to be $1 M_{\odot}.$ In order to perform this calculation we
place $N_0=10^6$ particles  
around the $L_5$ point, $x_0\in[-0.98;-0.88],\;y_0\in[-0.2;0.2].$ 


\begin{figure}
	\includegraphics[width=80mm]{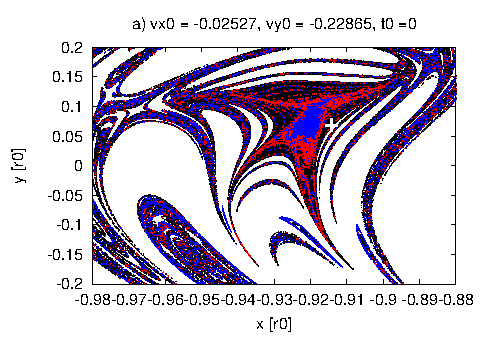}\\
	\includegraphics[width=80mm]{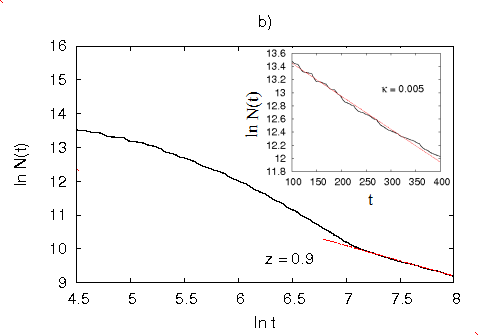}
    \caption{RFBP. (a) Initial positions of non-escaped trajectories for $T=$400, 1300, and 13000 days, black, red, and blue dots, respectively. The mass of the Sun is 1 solar mass, the initial velocities are the same as in Fig.~\ref{fig:rtbpl5} and the white cross  again shows the position of the $L_5$ point in RTBP. The filamentary fractal structure indicates the complex dynamics. (b) Number of surviving particles in time. The log-lin inset indicates the exponential decay corresponding to fast escape with rate $\kappa\approx 0.005$ while the linear fit to the tale of the log-log plot shows the algebraic decay for longer times, $z\approx 0.9.$}
    \label{fig:diffesctime}
\end{figure}


\begin{figure*}\includegraphics[width=55mm]{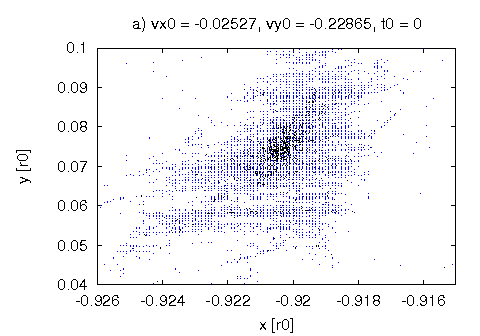}
\includegraphics[width=55mm]{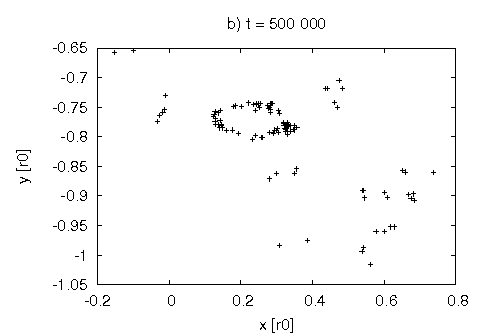}
\includegraphics[width=55mm]{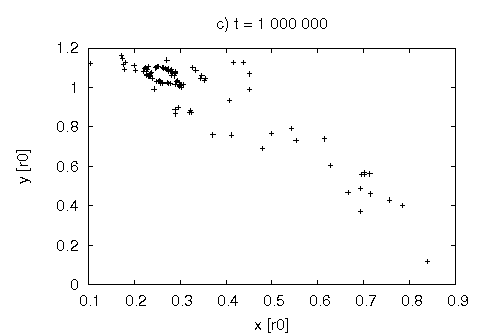}
\vspace{0.5cm}
\caption{RFBP. Existing for a very long time. (a) Blue dots draw the same pattern as in Fig.~\ref{fig:diffesctime} (a), non-escaped trajectories  for 13000 days. Black dots represent the initial conditions corresponding to particles that do not leave the system  for $10^6$ days. Panels (b) and (c) show that the very long-lived trajectories form a permanent structure in configuration space, indicating that the motion of these particles is regular.}
\label{fig:KAM_torus}
\end{figure*}

 The initial points of non-escaped trajectories for different integration times ($T=400,\;1300,\;13000$ days), are depicted by black, red and blue points, respectively  in
Fig.~\ref{fig:diffesctime} (a). The longer the integration time the
smaller the survival area. Consequently, we can identify a long-lived
island as a subset of \textit{shorter time} islands. Nevertheless, a
robust filamentary structure encloses the inner compact long-lived
island. Refined numerical investigations show that the filamentary structure has a self-similar nature 
and, therefore, one might expect chaotic behaviour
in this part of the phase space. 

It is known that in Hamiltonian dynamics the number of non-escaped
trajectories shows an exponential decay for shorter times and follows
a power-law for longer times ({\it see for example Lai \& T\'el (2011)}). Short
time escapes are related to the transversal intersections of the stable and
unstable manifolds of the UPOs where the dynamics can be considered fully 
hyperbolic. Particles with longer life time are confined to that
region of phase space where the manifolds intersect tangentially. 
This phenomenon appears, in fact, close to the border of regular tori
and is known as stickiness. The crossover from exponential to power law decay can 
indeed be seen in Fig.~\ref{fig:diffesctime} (b). 

 Fig.~\ref{fig:diffesctime} (b) shows the two
segments. The straight line in the inset on semi-logarithmic scale
indicates the exponential decay  ($N(t)\sim e^{-{\kappa}t}$), while the linear fit provides for escape rate
$\kappa\approx0.005.$ One can estimate (Altmann, Portela \& T\'el 2013)  the average lifetime of chaos
($\tau$) from the escape rate, $\tau\approx 1/\kappa\approx200$ days. 
The exponent $z$ in
algebraic decay ($N(t)\sim t^{-z}$) is found to be $z\approx 0.9$
which is somewhat smaller than the actual accepted, probably
universal, value ($\sim 1.5$) of chaotic transport near the remnants of
the last KAM torus (Meiss 2015).

To find analytically, and also numerically, the unstable periodic orbits
and their manifolds in high dimensional phase space is a challenging
task. The  motion in a $2N$ dimensional phase space ($N$ is
the number of degrees of freedom) proceeds on a $2N-M$ dimensional
surface where $M$ is the number of existing first
integrals. Constructing any Poincar\'e surface of section in $2N-M$ 
dimensions, if $2N-M>3,$ yields a confused structure-less picture, since we see
the projection of any remaining dimensions.

Considering that in our  model the phase space has 12
dimensions, we are not 
able to construct and visualize a Poincar\'e surface. More precisely,
any 2D section of the phase space has a fuzzy structure which actually shows
the projection of all the rest components of the positions
and velocities. In contrast, the above defined ICMs
are suitable to visualize the escape dynamics even on Poincar\'e surfaces
of sections.  This means, the initial conditions in general belong to a
well-defined configuration of the system, i.e. Moon's and Sun's
positions and all the velocities are fixed. Therefore, the escape dynamics are
fully characterized by the initial coordinates and velocities of the test
particle for a given energy level. Using a statistical approach by
means of an ensemble, the initial positions in $x-y$
configuration plane depict the filamentary fractal set corresponding
to the finite time chaotic motion before escape. This object can be
thought as of the intersection of the $x-y$ plane and the stable
manifold of a higher dimensional chaotic saddle.
Therefore, we want
to use the ICMs to apply the control to the spacecraft's motion.

Interestingly, however, one might also find regular structures in the
configuration space. Very long calculations represent that there are
trajectories that form a torus in the $x-y$ plane, see
Fig.~\ref{fig:KAM_torus}. These points represent test particles orbiting
the Earth at $t=5\cdot 10^5$ days, panel (b), and
$t=10^6$ days panel (c), respectively. Higher dimensional ($D>2$)
tori can actually be identified and characterized according to various
analytical and numerical methods
(Jorba 2000; Lan, Chandre \& Cvitanovi{\'c} 2006; Laakso \& Kaasalainen 2014; Lange et al. 2014).
As we already mentioned above,  objects in the
phase space, like the torus in Fig.~\ref{fig:KAM_torus}, associated to
regular motions are responsible for algebraic decay of very long-lived
trajectories.

\subsection{The Effect of the Initial Velocity}

So far in all our investigations,  the test particles started with the same velocity ($v_{x0}$; $v_{y0}$)
which is equal to the velocity of the $L_5$ point in the circular RTBP, but it may be interesting to examine what happens to the long- lived island around the $L_5$ point if  we change the initial velocity.

\begin{figure*}

\includegraphics[width=55mm]{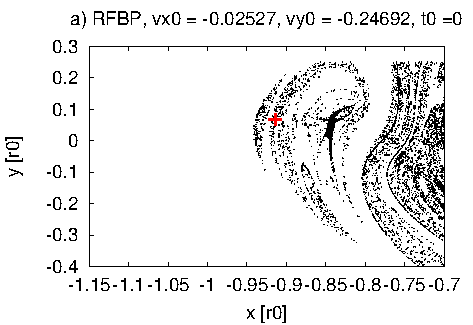}
\includegraphics[width=55mm]{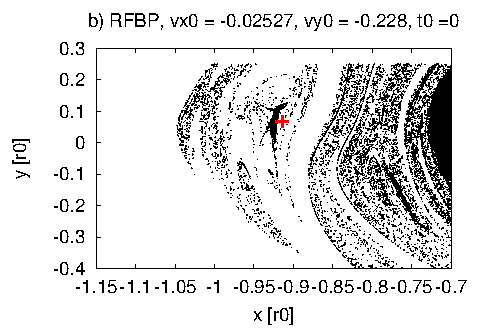}
\includegraphics[width=55mm]{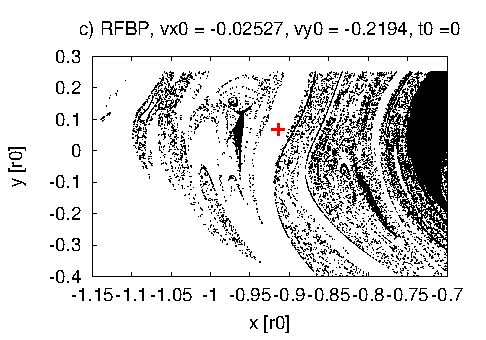}
\vspace{0.5cm}
\caption{RFBP. Initial condition maps ($T$=1300) for different
  velocities. The initial velocity component $v_{x0}$ is equal to the
  velocity of the $L_5$ point while $v_{y0}$ in (a) larger, (b) equal ,
  and (c) smaller  than the $v_{y0}$ of $L_5$. While decreasing the initial
  velocity, the long-lived island drifts away from the Earth. The panels show ICMs for 360 000 test particles. The red cross denotes the position of   $L_5$ point in RTBP.}
\label{fig:velocity}
\end{figure*}

In this subsection we investigate how the structure of the ICMs changes for
different initial velocities. To do this, several ICMs are
generated with given conditions. With this procedure one can scan the $x-y$ plane for different
$v_{y0}$ keeping $v_{x0}$ constant. Fig.~\ref{fig:velocity}
shows three ICMs ($T$=1300 days) for different velocities $v_{y0}.$ 

Panels (a), (b), and (c) show
ICMs where initial conditions are chosen from the region
$x\in[-1.15,-0.7],\;y\in[-0.4,0.2]$ including the exact position of $L_5$
point in the RTBP (red cross). $v_{x0}=-0.02527$ in all 
three cases ($v_{x0}$ is the initial velocity component of the $L_5$
point) while $v_{y0}=-0.24692\;-0.228,\;-0.2194$ for panel (a), (b),
and (c), respectively. Different initial $v_{y0}$ velocities
correspond to  smaller, equal, and larger values than that of the $L_5$ point's $v_{y0}.$
One can clearly see that, depending on  the initial
velocity,  the relative position of the long-lived island compared to
the red cross changes. The smaller the $v_{y0}$,  the further the
long-lived island  from the origin. 

We can point out that small changes in $v_{y0}$ results  an
outward/inward drift of the structures in $x-y$ plane. This fact will be
the base of our control method, as we shall see later.  

\subsection{How to Keep a Spacecraft in a Confined Region}

In this section we present our strategy  on how to keep a test particle in the
system, i.e. preventing the scenario in which $r$ exceeds 1.5. Though
the orbit is well-defined by the initial conditions, 
our method does not allow a strict station keeping,  but a rather cheap
control of the motion that allows for  a bounded orbit around the initial
one. 

\begin{figure*}
\includegraphics[width=55mm]{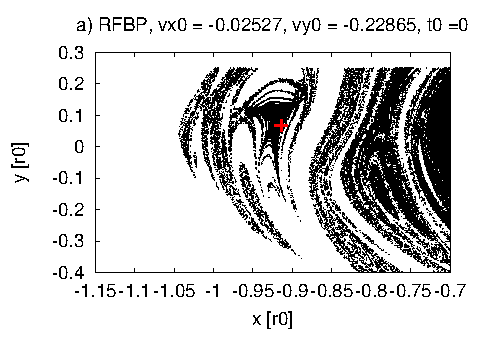}
\includegraphics[width=55mm]{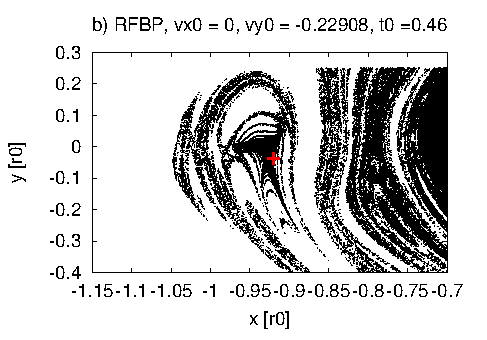}
\includegraphics[width=55mm]{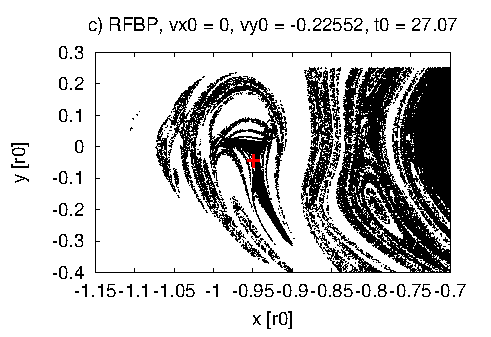}\\
\includegraphics[width=55mm]{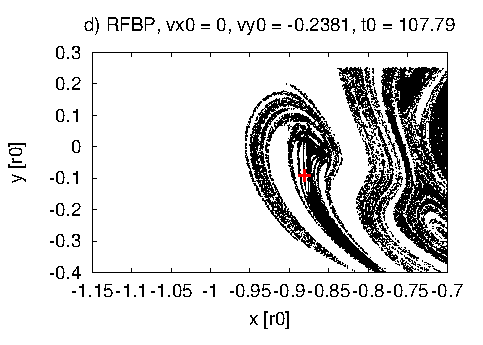}
\includegraphics[width=55mm]{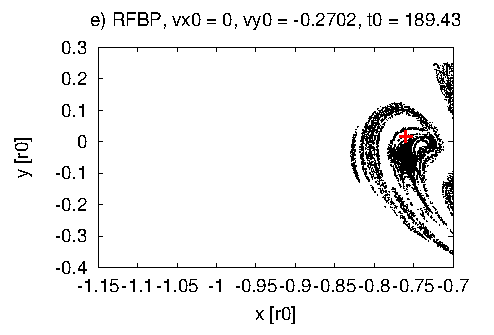}
\includegraphics[width=55mm]{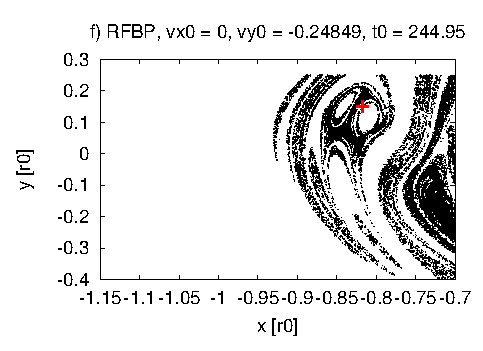}\\
\vspace{0.5cm}
\caption{RFBP. Initial condition maps ($T$=400 days) at different times. Panel (a) shows the ICM associated to the starting  $t_{0}=0,\;v_{x0}=-0.0228$ (0. ICM). ICMs in other panels correspond to the condition $v_{x0}=0,\;v_{y0}<0$ related to the RT, (b) $t$=0.46 day (1st), (c) $t$=27.07 days (2nd), (d) $t$=107.79 days  (5th), (e) $t$=189.43 days  (7th), and (f) $t$=244.95 days  (10th). The RT is marked by  a red cross. No control is applied during the motion. Note that not all maps are presented between $t$=0 and $t$=244.95 days.}
\label{fig:no_control}
\end{figure*}

The basic idea of the method is to construct ICMs
for appropriate initial velocities similar to that what we have seen in
Fig.~\ref{fig:velocity}. Fig.~\ref{fig:no_control} (a) shows the ICM
containing $3.6\cdot10^5$ test particles  starting from the region 
$x_0\in[-1.15,-0.7],\;y_0\in[-0.4,0.3]$ with the velocity of the $L_5$ Lagrangian
point. The RT  is selected in such a way that
$(x_0,y_0)=(-0.914,0.068)$ and $(v_{x_0},
v_{y_0})=(-0.02527,-0.22865),$ see Table~\ref{table:epoch} third
column. These initial  
conditions correspond to the $L_5$ point in RTBP. Test integrations show
that the RT leaves the system  at  around 500 days. Therefore, the task is
to keep this trajectory in the system for  an  arbitrary  long period of time ($t\gg\tau$). The ICM is
constructed as usual,  where black dots represent the trajectories not
leaving the system until $T$=400 days,  and the red cross shows the RT on
the map. 

The next step is to follow the RT until the criterion $v_x=0 , v_y<0$  is
fulfilled. Then we check the actual $v_y=-0.229$ of the RT, and
depict a new ICM again for $T$=400 days, with these velocities
($v_{x0}=0 , v_{y0}=-0.229$) as updated  initial velocities, see 
Fig~\ref{fig:no_control} (b).

Since the initial velocity $v_{x_0}$ of the $L_5$ point is very
close to zero, we do not have to wait too long for the second ICM.
As indicated in the above  panel (b) 0.46 days were spent between  the 1st and
2nd ICMs and, therefore, the structure of panel (b) is very similar to the 
initial one, but some distortion (shift and skew) can be observed. 
It can be noticed that the red cross has left the contiguous
domain and is situated exactly on a nearby filament.

\begin{figure*}
\includegraphics[width=70mm]{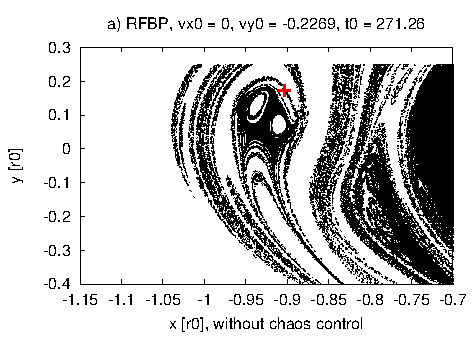}
\includegraphics[width=70mm]{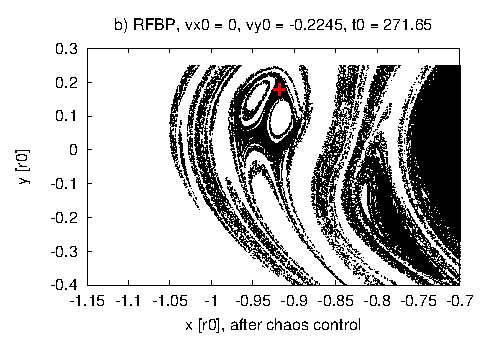}
\includegraphics[width=70mm]{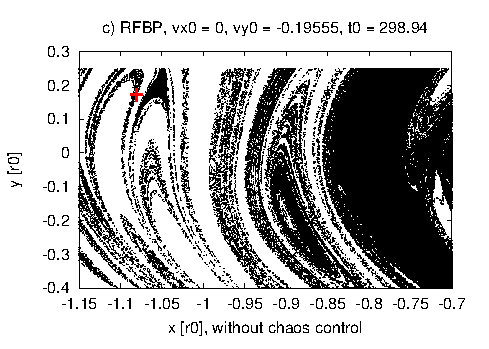}
\includegraphics[width=70mm]{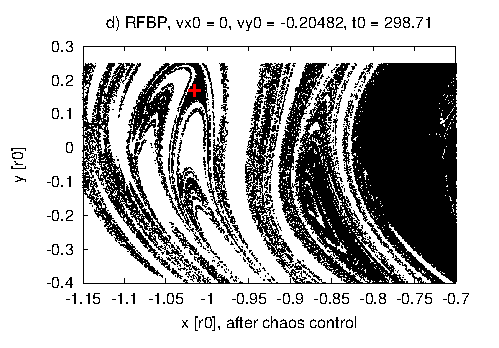}
\includegraphics[width=70mm]{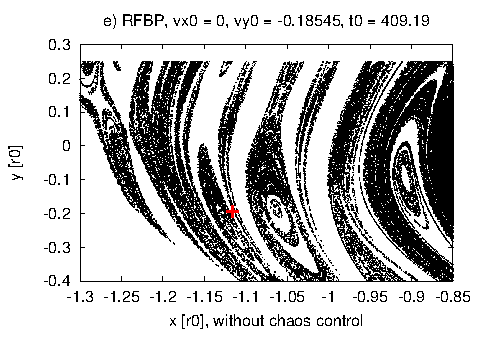}
\includegraphics[width=70mm]{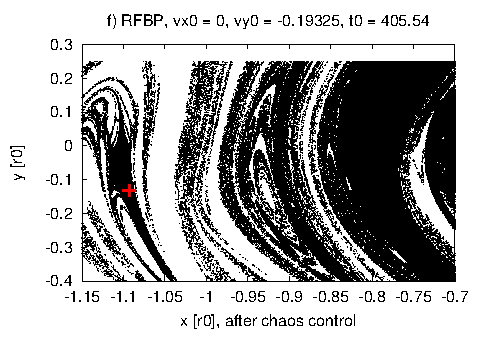}
\includegraphics[width=70mm]{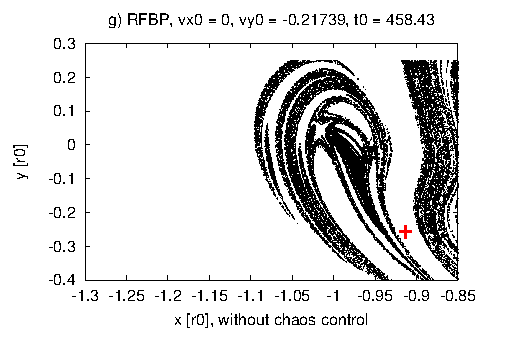}
\includegraphics[width=70mm]{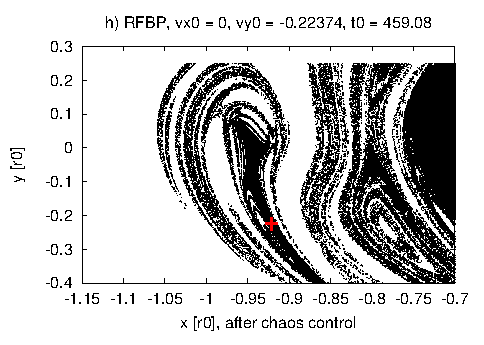}
\vspace{0.5cm}
\caption{ICMs ($T$=400 days) showing the effect of the
  intervention. The left column (a,c,e,g) depicts a continuation of
  Fig.~\ref{fig:no_control} forward in time $t$=271.26--458.43 days
  (11,12,16,18th). Not all ICMs are shown. The red cross denotes the
  position of the RT without control.  The right column (b,d,f,h) shows the
  ICMs (11,12,16,18th) after the intervention is performed. A slightly different structure of individual maps appears at different times, $t$=271.65--459.08 days. One can observe that the RT, red cross, remains inside the black long-lived island until $t\approx$460 days.}
\label{fig:control}
\end{figure*}

Now we repeat the procedure until the RT crosses the escape border $r=1.5$ at
roughly $t\approx$ 460 days. This will result in a series of ICMs for
$T=400$ days with different $v_{y_0}$ and zero horizontal velocity. Some of these maps are shown in
Figure~\ref{fig:no_control}. As one would expect, the consecutive ICMs follow each other with fairly constant times which 
related to the Moon's orbital period, $\sim 27$ days. The
corresponding epochs of various ICMs are shown above the panels. 

Nevertheless, there is a more robust and interesting phenomenon  that
can be observed if we look at carefully the different ICMs. Namely, the
RT moves along the filamentary structure until its escape. At the
beginning of the integration, as we have seen, the RT is on the filaments very
close to the long-lived island (Fig.~\ref{fig:no_control}, panels (a) and (b)).
Then it moves away, but still along the filaments $250\lesssim t\lesssim 450,$ 
Fig.~\ref{fig:control} left column (panels a, c, and e).
Finally the red cross lies in the white region which corresponds to an extremely 
fast escape, see  Fig.~\ref{fig:control} (g). 

The RT escapes without any intervention after $t\approx$ 450 days, which is on the same magnitude as
the average lifetime ($\tau$) introduced in Section 3.3. Having constructed all the ICMs until the escape 
(Fig.~\ref{fig:no_control} and Fig.~\ref{fig:control} left column), we can formulate the basic
objective of the control procedure. That is, \textit{the RT
should be shifted from filaments to a long-lived island with the  lowest possible energy
consumption.}


\begin{figure*}
\includegraphics[width=84mm]{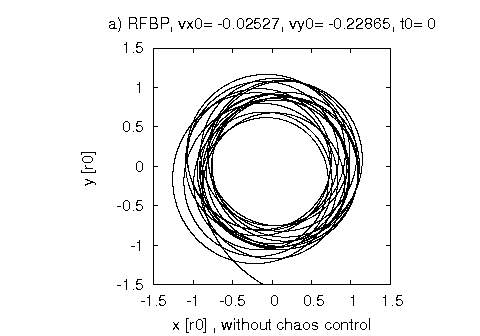}
\includegraphics[width=84mm]{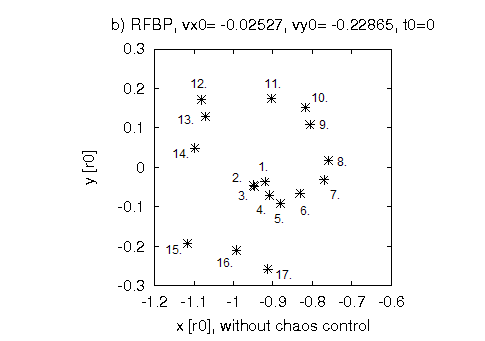}\\
\includegraphics[width=84mm]{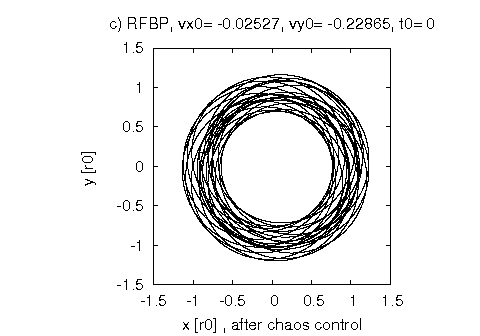}
\includegraphics[width=84mm]{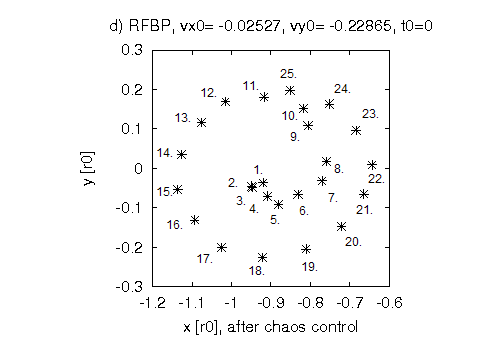}
\vspace{0.5cm}
\caption{Panels (a) and (b) show the uncontrolled, while panels (c) and (d) display the controlled motion. The corresponding section points of the RT with $v_x=0,\;v_{y}<0$ depicted in $x-y$ plane in panels (b)  and (d). Numbers beside the points correspond to the serial numbers of ICMs in Figure~\ref{fig:no_control} and \ref{fig:control}. The control procedure applied between  the 10th and 11th ICMs results in an ordered structure in $x-y$ plane which remains for  a very long time ($t$>30000 days).}
\label{fig:traj}
\end{figure*}

There are two essential questions arising: 1. How to do this?
2. When to do this? 

We start with the answer to the second question. Obviously, the control must start
before the RT leaves the filamentary structure. A good indicator of
this is the average lifetime of chaotic behaviour, $\tau\approx$ 200
days. Let us consider the ICM corresponding to instance $t=$244.95, which
is the 10th ICM after $t$=0, Fig.~\ref{fig:no_control} (f). We might
want to start the control at this time and see whether the red cross
lands in the long-lived island in the next ICM, at $t=$271.26 days
(Fig.~\ref{fig:control} a). 

 At this point we should address the answer to the first question. Since
we know the system dynamics only at discrete times, at every single
ICM, a natural choice
is a continuous tangential force acting on the RT (or
spacecraft/satellite) between two ICMs in the following form,
\begin{equation}
\mathbf{f} =  \left\{ \begin{array}{ll}
         F_y\cdot\mathbf{v_y}/|v| & \mbox{if $\;t \in \left[t_{n},\, t_{n} + \Delta t\right]$}, \\
         F_x\cdot\mathbf{v_x}/|v| & \mbox{if $\;t \in \left[t_{n},\, t_{n} + \Delta t\right]$}, \\
         0   & \mbox{otherwise},\end{array}\right.
\label{eq:force}
\end{equation}
where $F_x$ and $F_y$ are the magnitude of the applied force in directions  $\mathbf{v_x}$ and  $\mathbf{v_y}$  (where $\mathbf{v_x}$ is the $x$ component,  $\mathbf{v_y}$ is the $y$ component of the velocity vector), $t_{n}$ is the instance when the control begins and
$\Delta t$ describes the length of the control.  

In general, we switch on the propulsion at $t$ (or equivalently when we reach the criterion
$v_x=$0 and $v_y<$0)  in order to manoeuvre the spacecraft to a given phase space
position, $(x,y,v_x,v_y).$ Therefore, we should have an approximate initial guess for the
magnitude of the force (thrust or breaking). 
The energy conservation between two consecutive ICMs can serve a valuable choice for the force
we need. (See the Appendix \ref{app:1} for details.)

However, there is a problem if we adjust the velocity of the
satellite. Namely, the next ICM in the row will be at a different time
with a slightly different structure than it should be without
control.   

\begin{figure*}
\includegraphics[width=84mm]{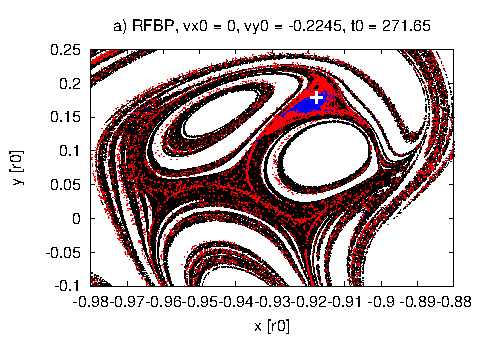}
\includegraphics[width=84mm]{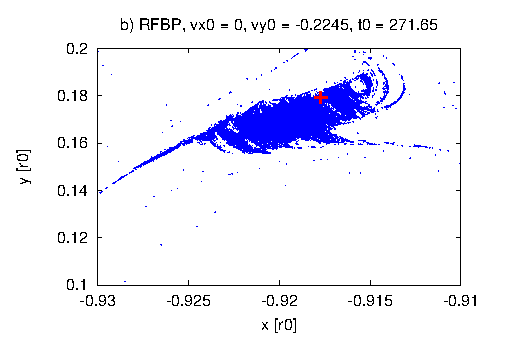}\\
\includegraphics[width=84mm]{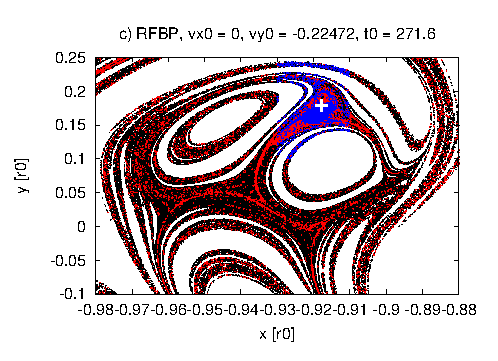}
\includegraphics[width=84mm]{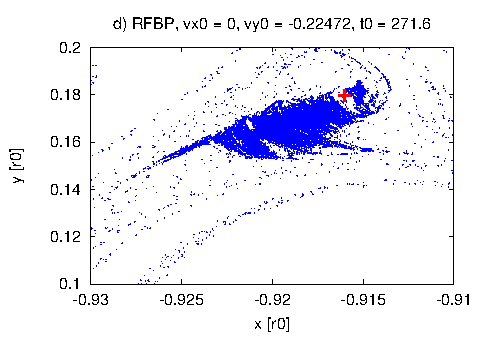}
\vspace{0.5cm}
\caption{ICMs for different integration times, $T$=400, 1300, 13000
  days. The longer the integration time, the smaller the size of the long-lived island (black, red, and blue regions, respectively), panels (a) and (c). Different control forces result different ICMs, i.e. the condition $v_{x0}=0,\;v_{y0}<0$ occurs at slightly different times and, consequently, different structures appear on the ICMs. Due to the different force the position of the RT will also be different. Therefore a fine tuning of the control can be achieved by pushing the RT into a very small but very long-lived island (b). With a different force, this fortunate place can be kept off (d). The cross shows the position of the RT.}
\label{fig:diff_force}
\end{figure*}

Let us consider a particular exercise to provide a better insight into the
control. Fig.~\ref{fig:no_control} and the left column of 
Fig.~\ref{fig:control}  (panels a,c,e, and g) show what happens if no control is
used. In brief, the red cross leaves the
pre-defined ($r$<1.5) region after $\sim$ 450 days. As we know from
previous simulations, the average lifetime of the chaotic behaviour
before escape is roughly 200 days. So, we should try to apply the
control procedure around this time. For instance, we can choose the 10th
ICM at $t=$244.95 (Fig.~\ref{fig:no_control} f) to start the procedure,
and then we want to reach the long-lived island appearing on the next
ICM, 11th in the row at $t=$271.26, Fig.~\ref{fig:control}(a). However, as  discussed earlier,
this ICM will have a different structure at different time ($t=$271.65), due to the intervention applied
between two maps, compare Fig.~\ref{fig:control} (a) and
(b).  Equation~(\ref{eq:Denergy}) provides an approximate 
value to the tangential force that should be used for the control,
$F=\sqrt{{ F_x}^2 +{F_y}^2}\approx 2.1\cdot 10^{-6}$ N. This force is valid in cases  the
satellite has a mass of 1 kg. This operation is performed between
$t_{10}$ and $t_{11}$ for $\Delta t=$26.9 days, see Eq.~(\ref{eq:force}). The sign of the force, i.e. thrusting or
pushing back, can be obtained from the relative position of the
RT and the long-lived island. For example, if we want to push the RT (red
cross in Fig.~\ref{fig:control}(a)) slightly outside, then we need a
push back force. The smaller the velocity, the larger the distance to
the central object. Table~\ref{table:control} summarizes the result of
the control procedure. The first row contains the coordinates of the RT,
belonging to the 10th ICM $v_x=0, v_y<0$ section point and time. The second row shows the coordinates of the RT at the next, the 11th section point without chaos control. In the third row we can see the coordinates of the RT at the 11th section point after chaos control between the 10th and the 11th section points.

\begin{table*}
\caption{Coordinates and velocities of the RT in the 10th and 11th ICMs with and without control.}
\begin{tabular}{llllll}
\hline
& $x [r_0]$ & $v_x [ r_0/day]$ & $y [ r_0]$ & $v_y [ r_0/day]$ & time of the ICM [day]\\\hline
& -0.81724699846 & 0  & 0.15185082461  & -0.24849685052 & 244.95
                                                                       (before
                                                                       control, 
                                                                       Fig.~\ref{fig:no_control}
                                                                       (f))\\
& -0.90315303819 & 0   & 0.17493609008  & -0.22690346315 &
                                                                       271.26 (without control, Fig.~\ref{fig:control}
                                                                       (a)) \\
& -0.91770366237 &  0 & 0.18092762375 & -0.22450011830 &
                                                                     271.65 (after control, Fig.~\ref{fig:control}
                                                                       (b))\\ \hline    
\end{tabular}
 \label{table:control}
\end{table*}

In fact, the approximation of the control force does not guarantee
that we can reach the selected domain of the phase space. In this
situation a successive iteration is needed to drive the trajectory
precisely to the appropriate place. In our case, a factor of $\sim$2 is needed
to have a perfect force, $F\approx4.86\cdot 10^{-6}$ N. If the manoeuvre
is successful (e.g. Fig.~\ref{fig:control} b), which means the RT
pinpointed the desired long-lived island, we can be sure that the
trajectory will not escape, at least during the next 400 days, i.e. it
remains in the black long-lived region, see the 
right column of Fig.~\ref{fig:control}, panels( b), (d), (f) and (h). And if, after a while, it leaves that domain along a filament,
just a simple repetition of the above control procedure is needed to
keep the particle in the system at least for $\approx \tau$ time. 

Fig.~\ref{fig:traj} depicts the result of the control considering
the RT. The first row illustrates the uncontrolled
motion with an escape after several
revolutions in the $x-y$ plane. While the second row
shows the motion after the control, we
refer to Fig.~\ref{fig:control} (a) and (b) and Table~\ref{table:control}.  
We also note that Fig.~\ref{fig:traj} (b) and (d) correspond to a section of the phase
space defined by conditions $v_x$=0 and $v_y<$ 0 similarly to ICMs but
show all the consecutive section points of RT (marked by numbers).
It is clearly seen that after the intervention, the section points are 
aligned around a "smooth" spiral indicating that the motion becomes regular in
some sense.

However, as
we have already seen in Fig.~\ref{fig:rtbpl5}, there are subsets
of a certain ICM corresponding to longer lifetimes. That is, if we know
the position of such a sub-region, the control force can be fine tuned
in order to land in this smaller but long-lived island. To visualize
this we present two  ICMs for $T$=400, 1300, and 13000 days at
slightly different epochs, see
Fig.~\ref{fig:diff_force}. The first row shows the ICM with the force ($F_x = F_y = 4.89\cdot
10^{-6}$ N) that we used during the control process. This results in a
trajectory bounded for more than 36000 days, Fig.~\ref{fig:traj} (a) and
(b). Fig.~\ref{fig:diff_force} (b) indicates that the RT lies in the
domain of longer life time. In contrast, using a somewhat smaller
force , $F_x=F_y=4.29\cdot10^{-6}$ N, we reach the ICM 0.05 days earlier, 
at t=271.6 days. In this map (Fig.~\ref{fig:diff_force} d) the
destination of the RT is outside the blue region, consequently,
the RT leaves the system earlier. 

In addition, we found that the energy consumption in case of a
spacecraft with a mass of 1 kg is 11.3 kJ, while in a more realistic
case, say, the SOHO satellite, which has a mass 1850kg, $\Delta E\sim$
20.9 MJ. 

\subsection{Transfer from Low Earth Orbit (LEO) to Low Lunar Orbit (LLO) through the $L_5$ Point }

In the following we show a transfer from low Earth orbit (LEO: 189 km 
circular orbit above the ground) to a
circular low Lunar orbit (LLO: 106 km above the Moon's surface) through the $L_5$ point's long-lived island.

In this subsection we demonstrate how to use the above mentioned chaos
control method for low energy interplanetary transfers between Earth
and the Moon. If the dynamics are regular, i.e a two-body approach is valid,
large corrections cannot be achieved by small parameter changes and,
therefore, no significant energy savings can be made compared to the
classical methods (Hohmann transfer, weak stability boundary method). In
other words, the method works only in the chaotic phase space regime,
(far from the Earth and Moon) where unstable dynamics govern the
motion under the gravitational influence of both objects.

\begin{figure*}
\includegraphics[width=55mm]{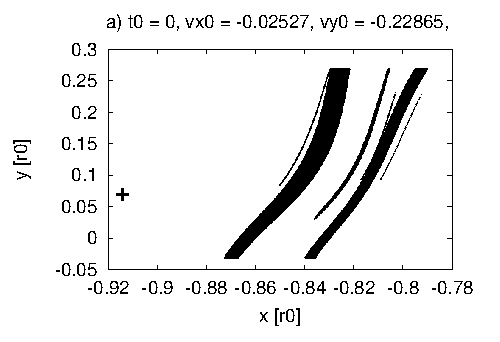}
\includegraphics[width=55mm]{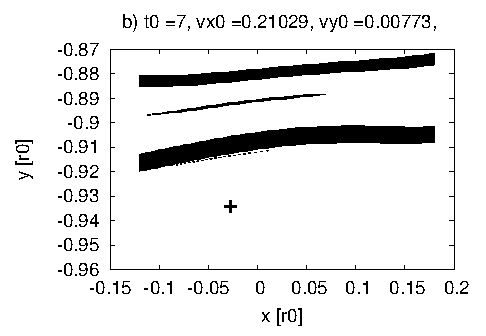}
\includegraphics[width=55mm]{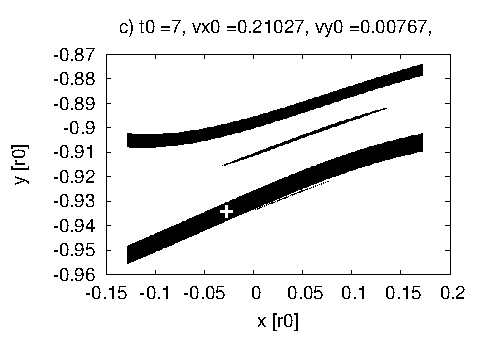}
\vspace{0.5cm}
\caption{ICMs for 1 000 000 trajectories. The starting region is the same
  as in Fig~\ref{fig:coosys}. Black stripes represent the initial
  conditions leading to low Lunar orbits ($\sim$100 km to the Moon).
The initial velocity is (a) equal to the velocity of the $L_5$
point. Panels (b) and (c) depict the initial condition maps after 7 days
without and with the control. The  cross shows the reference trajectory.}
\label{fig:target}
\end{figure*}

Therefore, one needs to reach that domain of phase space where
unstable dynamics dominates, and control can be easily realized. A good
place to do this is the theoretical position
of the Lagrangian $L_5$ point of the Earth-Moon-satellite RTB system. As
shown before, the $L_5$ point is not stable anymore if the Sun's
perturbation is taken into account but (arbitrarily) long-lived islands can be found
that are suitable for control and temporal parking the space craft at
the Moon's orbit. Moreover, several initial conditions near the $L_5$
point lead to a Moon crashing orbit as we will see later in Fig.~\ref{fig:target}.

We suppose an ion propulsion thrust. That is, a continuous acting low
fuel-cost manoeuvre instead of an impulse like force. Our calculations are  compared to the widely used
weak stability boundary (WSB) method published  {\it  by Belbruno \& Miller
  (1993)}. Similar to the WSB method we focus only on the transfer
between LEO and LLO and ignore the energy cost of launch from the
Earth basis and the breaking at vicinity of the Moon.

\begin{figure}
\includegraphics[width=80mm]{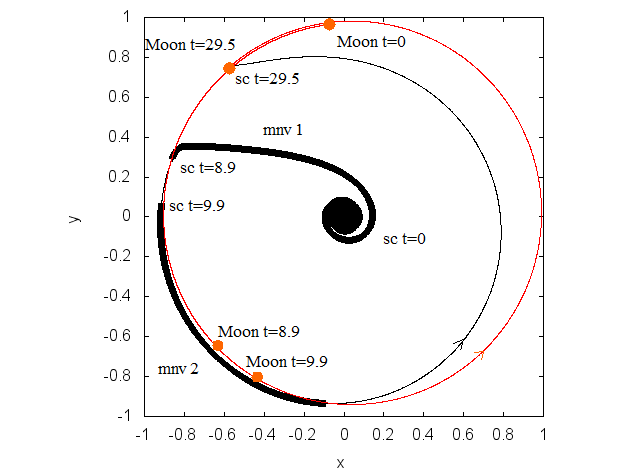}
\vspace{0.5cm}
\caption{Flight from a low Earth orbit (198 km above the
  Earth's surface) to a low lunar
  orbit (106 km above the Moon's surface) through the $L_5$ long-lived island.  Bold lines
  represent the orbit segments under control (mnv Nr. 1,2,3). The
  whole mission takes roughly 1 month and the energy consumption is
  compatible  with other transfer methods. The
  corresponding data can be found in Table~\ref{table:orbits}. }
\label{fig:orb}
\end{figure}

The first stage of our transfer consists of the orbit from LEO to a
long-lived island around the $L_5$ point, similar to the ballistic capture
method (Salazar, Macau \& Winter 2014). Due to the appropriate continuous thrust
($F\sim v_y$), the spacecraft will spiral out from LEO until reaching the escape velocity and
leaving the Earth's range in an elongated elliptical orbit which has
a turning point (apogeum) at the Moon's orbit, Figure~\ref{fig:orb}. This part of the
manoeuvre is similar to the Hohmann transfer, with a difference that
the spacecraft arrives at the Moon's orbit exactly at the $L_5$ point,
instead at the Moon itself. This fact allows us either to keep the
satellite in the long-lived island with the above mentioned chaos
control method or to eject it directly to the Moon after a while. 

This can  later be done by a small control force which enables  finding 
the initial conditions that lead to an LLO. Figure~\ref{fig:target}
shows the initial condition map in the vicinity of $L_5$, similar to
Fig.~\ref{fig:rtbpl5}. The cross shows the reference trajectory
after the first manoeuvre, while the stripes denote the desired (x,y)
co-ordinates in order to get LLO. Using a breaking force for
$t=6.68$ days, the reference trajectory can be pushed into the
bottom stripe (Figure~\ref{fig:target}c), and after 29.54 days it will
reach the LLO. In detail, by fine tuning the thrust we can get a reference
trajectory that approaches the Moon at 106 km (cross in Fig.~\ref{fig:target}). The required 
energy is 58.38 kJ for a 1 kg space craft, which  corresponds to
$\Delta$V = 0.0156  km/s. This intervention is the second
control manoeuvre in Fig~\ref{fig:orb}. After this control we need
29.54 days to reach the LLO.

Table~\ref{table:orbits} gives the summary of the controls applied during the
entire mission. The energy consumption of the first manoeuvre, from the LEO to the
$L_5$-long-lived island, is 29.31 MJ. This amount of energy is equals to
that of the Hohmann transfer cost of  29.32 MJoule ($\Delta V_1$ = 3.132 km/s).


 Manoeuvre 2 from $L_5$-orbit to the low Lunar orbit needs a velocity adjustment of 
$\Delta V_2 $= 0.6933 km/s (0.058MJ, see above). Comparing this result to the data of  
Table~3 (mnv. 2) {\it in Belbruno \& Miller (1993),} we find that this is only
slightly less than the desired $\Delta V=$ 0.695 km/s in weak stability
boundary method, but the time of flight is
much less, $\sim$1 month instead of 3-5 months (Belbruno \& Miller 1993).

Manouevre 3 is the velocity reduction to the circular Lunar orbit at
an altitude of 106 km above the Moon's surface. Table~3 contains only
the required $\Delta V$ since the time of intervention can be chosen
arbitrarily.


\begin{table*}
\caption{Data of the transfer from low Earth circular to low lunar circular orbit. $t_i$ signs the beginning of the $i^{th}$ manoeuvre  ($i= 1,2,3$ respectively), $\Delta$t is the duration of the control,  $\epsilon$ is the invested energy. }
\begin{tabular}{llllllllll}
\hline
& manoeuvre & $t_i$ [day]& $\Delta$t [day]&  $F_x$ [N] &$ F_y$ [N] &   from & to  & $\epsilon $ [MJ] & $\Delta$V [km/s]\\\hline
& 1. & 0  & 8.9 & 0 & 0.0258 &  LEO & $L_5$ orbit & 29.3 & \\
& 2. & 9.9   & 6.68 & -0.0001 &-0.0001 & $L_5$ orbit &to the Moon  & 0.05838  & 0.0156  \\
& 3.  &  29.54 &  -- & --  & -- & 106 km above the  Moon & LLO  &   &  0.6777 \\ \hline    
\end{tabular}
 \label{table:orbits}
\end{table*}

\section{Summary and Conclusions}

 We published our first results  in {\it Sl\'iz , S\"uli \& Kov\'acs (2015)}, where we presented our chaos control method  for one special case:  to keep a spacecraft (as a 1-kg test particle) near the Earth-Moon $L_5$ point  in the presence of the Sun (RFBP).
Since then we realized that this method is generalisable,  and also suitable for targeting a spacecraft. In this work   we presented a generally applicable method to keep a spacecraft  in a confined region or targeting it, when on the ICM there is a  long-lived island  in the "nearby". The applied force is very small, tangentially,  proportional to the velocity, and pushes the RT onto this island. We presented two examples: 1. how to keep a spacecraft  in a pre-defined region, and 2. how to go to the Moon (from LEO to LLO). In both cases the applied force is  within the range of the ion thruster. We showed the influence of the Sun's mass, the integration time and the initial velocity  to the size,  shift  and  shape of the long-lived islands around the Earth-Moon $L_5$ point. Regarding the shift of  $L_5$ point for different initial velocities, we found the same result as  {\it Schwarz \& Eggl (2011)} found in  binary systems.  

Numerical simulations show that the presence of the Sun has dramatic effects on  the size and the position of bounded motion around the triangular equilibrium point $L_5$. Hence, we pointed out that the question of keeping a particle in the system is relevant, even if we start the RT very close to the long-lived island near the Lagrangian point $L_5$, which is thought to be linearly stable in the restricted three body problem. 

Using the methodology of transient chaos, we can show that the filamentary structure of ICMs are related to the stable manifold of the so called chaotic saddle. Moreover, the escape dynamics are found to be a classical phenomenon of the finite time chaotic behaviour and therefore, the well-known statistical description for non-escaping trajectories can be used. That is, for short times an exponential decay of surviving trajectories can be seen. The dynamics of this regime are governed by the fractal set obtained around the long-lived islands in ICMs. On the other hand, the number of non-escaped trajectories shows an algebraic decay for longer times, which is a consequence of the stickiness effect around the KAM tori embedded in the phase space.

We have used a combination of two chaos control methods in order to have a suitable method of avoiding the escape of a test particle. The results show that, interestingly, the RT leaves the system along the filamentary structure constructed at different times and at different ($v_{y}$) velocities. Taking  advantage of this phenomenon, the essence of our control mechanism is pushing  the RT back  into a long-lived island whose future position is known from numerical integration.

Since we know the future dynamics of the satellite, this 
means, the structure of the subsequent ICMs are explored,
and we are able to choose different regions as a target domain,
depending on the survival time. In other words, applying a
slightly different force in the control process, the spacecraft can
land in that part of the phase space where the surviving time
is much longer than the average life time of the escape dynamics.
Consequently, the next intervention can be delayed by 
several thousands of orbits. Changing the force and the period
of the control in equation (1) can serve multiple options, such as wether 
to optimize the mission focusing either on the energy consumption
or the period of planned interventions.

The initial conditions we used could seem somewhat arbitrary.
We would like to note that the finite time chaotic
dynamics governed by the unstable periodic orbits and their
manifolds is a robust phenomenon in general. Therefore, an
obvious expectation is that different initial conditions provide
similar fractal structures  in ICMs than in the model we used.
Our preliminary results show that Keplerian initial conditions
of test particles led to a qualitatively similar filamentary
structure in phase space.

Finally, we should mention that dynamical control of a test body does
not  always mean  its prevention from escape. There might be situations
when escape is a more desirable scenario than keeping something close
to the Earth or Moon. Imagine we want to hijack an asteroid using the least amount of energy possible. 
A natural choice would be to push it away from the
filament into the white region where fast escape takes place
(Fig.~\ref{fig:no_control} and \ref{fig:control}). Our
simulations (not presented here) show that particles ejected from the
vicinity of the Moon's $L_5$ point terminated to near-Earth orbits
around the Sun. Initial condition maps and chaos control method
we used for our study  can also  be useful to model these scenarios,
either in planar or in spatial Sun-Earth-Moon-asteroid systems.
However, such computations are beyond the scope of recent work. 

 We also showed that the required energy for a transfer from
   LEO to LLO is about the same amount as the weak stability boundary
   method predicts, but the time of flight is significantly smaller.

The method described in this paper is realisable  with - even
  solar powered - electric  ion  propulsion producing very low
  (milli-newtons) thrust. It is true that in this case,  the journey requires a
  higher delta-v and usually there is a large increase in time compared to a
  chemical rocket, but the best efficiency 
  of electrical thrusters may significantly reduce the cost of the
  flight. In relation to the Earth-Moon, the increased
 flying time is not suitable for human
  space flight, but the situation is quite different in the case of interplanetary flight where the difference matters less. (Brophy \&  Rogers 2000).  

This chaos control method can be applied not only for celestial mechanics problems, but also  other fields of science, and can be used  even in engineering practice or biophysics, wherever behind  the chaotic phenomenon there is a time dependent  mathematic model. In celestial mechanics it can be applied  planning the mission to Mars or to other planets and/or to find an " interplanetary superhighway". 

\section*{Acknowledgements}

This work was partially supported by the Hungarian OTKA Grant Nr. NK 100296. The authors want to thank Dr. T. T\'el his valuable suggestions and helpful discussions.
  

%
%

\appendix

\appendix
\section{Estimation of the control force}\label{app:1}

Figure~\ref{fig:no_control} (f) and  Figure~\ref{fig:control}(a) indicate that the aim of a control is to achieve a desired small $\Delta r$ displacement in order to land into that part of the $x-y$ ICM, where a long-lived island is situated. It is also pointed out that if we perform a tiny but long control force, the structure of the next ICM depends on the force applied itself. Therefore, an initial guess as to the magnitude of the intervention is required. To have this information, we can use a simple calculation based on the energy conservation in two-body problem.
\begin{equation}
\Delta E=0.5m\Delta v^2-GM(\Delta r)^{-1}=\int F\mathrm{d}s,
\label{eq:Denergy}
\end{equation}
where $\Delta v^2=v_{n+1}^2-v_{n}^2$ and $\Delta r=r_{n}r_{n+1}/(r_{n}+r_{n+1})$ can be obtained from the $n$th and $(n+1)$th ICMs, $m$ and $M$ are the mass of the satellite and the Moon, respectively, $G$ denotes the Newtonian gravitational constant. In equation~(\ref{eq:Denergy}) index $n+1$ indicates the velocity and position of the desired point on the ICM, while index $n$ refers to the starting point on the preceding ICM. Applying equation~(\ref{eq:Denergy}) in case of Fig.~\ref{fig:no_control} (f) and Fig.~\ref{fig:control} (a), we get the estimation the absolute value to the control force $F\approx10^{-6}$N when the satellite has the mass 1kg and assuming a circular orbit. The later means that the r.h.s. of equation~(\ref{eq:Denergy}) is $F\cdot 2\pi r$ where $r$ denotes the Moon's orbital distance to the Earth while $F$ is the absolute value of the control force.

\end{document}